\pdfoutput=1
\documentclass[12pt,a4paper]{article}

\usepackage{color}

\usepackage{ifthen} 
\newboolean{pdflatex}
\setboolean{pdflatex}{true} 
\newboolean{articletitles}
\setboolean{articletitles}{true} 
\newboolean{uprightparticles}
\setboolean{uprightparticles}{true} 
\newboolean{inbibliography}
\setboolean{inbibliography}{false} 

\newboolean{prl}
\setboolean{prl}{False} 

\usepackage{lscape}
\usepackage{graphpap} 
\usepackage{multirow} 
\usepackage{rotating} 
\usepackage{mathrsfs}
\usepackage{mathtools} 
\usepackage{dashrule}
\usepackage{tikz}
\usepackage{enumerate}
\usetikzlibrary{patterns}
\usepackage{todonotes} 

\def\paperauthors{LHCb collaboration} 

\def\paperauthors{LHCb collaboration} 

\def\paperasciititle{Observation of new resonances in the Lambda_b pi+ pi- system} 
\def\papertitle{Observation of new resonances in the $\Lb\pip\pim$ system} 
\def\paperkeywords{{High Energy Physics}, {LHCb}} 
\def\papercopyright{\the\year\ CERN for the benefit of the LHCb collaboration} 
\def\paperlicence{CC-BY-4.0 licence}
\def\paperlicenceurl{https://creativecommons.org/licenses/by/4.0/}

\usepackage[top=1in, bottom=1.25in, left=1in, right=1in]{geometry}

\usepackage{microtype}
\usepackage{lineno}  
\usepackage{xspace} 
\usepackage{caption} 

\usepackage{graphicx}  
\usepackage{color}
\usepackage{colortbl}
\graphicspath{{./figs/}} 

\usepackage{amsmath} 
\usepackage{amssymb}
\usepackage{amsfonts}
\usepackage{upgreek} 

\newcommand*\patchAmsMathEnvironmentForLineno[1]{%
\expandafter\let\csname old#1\expandafter\endcsname\csname #1\endcsname
\expandafter\let\csname oldend#1\expandafter\endcsname\csname
end#1\endcsname
 \renewenvironment{#1}%
   {\linenomath\csname old#1\endcsname}%
   {\csname oldend#1\endcsname\endlinenomath}%
}
\newcommand*\patchBothAmsMathEnvironmentsForLineno[1]{%
  \patchAmsMathEnvironmentForLineno{#1}%
  \patchAmsMathEnvironmentForLineno{#1*}%
}
\AtBeginDocument{%
\patchBothAmsMathEnvironmentsForLineno{equation}%
\patchBothAmsMathEnvironmentsForLineno{align}%
\patchBothAmsMathEnvironmentsForLineno{flalign}%
\patchBothAmsMathEnvironmentsForLineno{alignat}%
\patchBothAmsMathEnvironmentsForLineno{gather}%
\patchBothAmsMathEnvironmentsForLineno{multline}%
\patchBothAmsMathEnvironmentsForLineno{eqnarray}%
}

\usepackage{hyperxmp}

\usepackage[pdftex,
            pdfauthor={\paperauthors},
            pdftitle={\paperasciititle},
            pdfkeywords={\paperkeywords},
            pdfcopyright={Copyright (C) \papercopyright},
            pdflicenseurl={\paperlicenceurl}]{hyperref}

\usepackage[all]{hypcap} 

\usepackage{xspace} 
\usepackage{upgreek}

\def\lhcb   {\mbox{LHCb}\xspace}

\def\cdf    {\mbox{CDF}\xspace}

\def\MagUp {\mbox{\em Mag\kern -0.05em Up}\xspace}

\ifthenelse{\boolean{uprightparticles}}%
{\def\Palpha      {\ensuremath{\upalpha}\xspace}

 \def\Pdelta      {\ensuremath{\updelta}\xspace}

 \def\Pmu         {\ensuremath{\upmu}\xspace}

 \def\Ppi         {\ensuremath{\uppi}\xspace}

 \def\Pchi        {\ensuremath{\upchi}\xspace}                 
 \def\Ppsi        {\ensuremath{\uppsi}\xspace}

 \def\PDelta      {\ensuremath{\Delta}\xspace}                 
 \def\PXi      {\ensuremath{\Xi}\xspace}                 
 \def\PLambda      {\ensuremath{\Lambda}\xspace}                 
 \def\PSigma      {\ensuremath{\Sigma}\xspace}                 
 \def\POmega      {\ensuremath{\Omega}\xspace}                 
 \def\PUpsilon      {\ensuremath{\Upsilon}\xspace}                 
 
 \def\PB      {\ensuremath{\mathrm{B}}\xspace}                 
                  
 \def\PD      {\ensuremath{\mathrm{D}}\xspace}

 \def\PJ      {\ensuremath{\mathrm{J}}\xspace}                 
 \def\PK      {\ensuremath{\mathrm{K}}\xspace}

 \def\Pb      {\ensuremath{\mathrm{b}}\xspace}                 
 \def\Pc      {\ensuremath{\mathrm{c}}\xspace}

 \def\Pi      {\ensuremath{\mathrm{i}}\xspace}

 \def\Pp      {\ensuremath{\mathrm{p}}\xspace}

}
{\def\Palpha      {\ensuremath{\alpha}\xspace}

 \def\Pdelta      {\ensuremath{\delta}\xspace}

 \def\Pmu         {\ensuremath{\mu}\xspace}

 \def\Ppi         {\ensuremath{\pi}\xspace}

 \def\Pchi        {\ensuremath{\chi}\xspace}                 
 \def\Ppsi        {\ensuremath{\psi}\xspace}                 
                  
 \mathchardef\PDelta="7101
 \mathchardef\PXi="7104
 \mathchardef\PLambda="7103
 \mathchardef\PSigma="7106
 \mathchardef\POmega="710A
 \mathchardef\PUpsilon="7107
                  
 \def\PB      {\ensuremath{B}\xspace}                 
                  
 \def\PD      {\ensuremath{D}\xspace}

 \def\PJ      {\ensuremath{J}\xspace}                 
 \def\PK      {\ensuremath{K}\xspace}

 \def\Pb      {\ensuremath{b}\xspace}                 
 \def\Pc      {\ensuremath{c}\xspace}

 \def\Pi      {\ensuremath{i}\xspace}

 \def\Pp      {\ensuremath{p}\xspace}

}

\makeatletter
\ifcase \@ptsize \relax
  \newcommand{\miniscule}{\@setfontsize\miniscule{4}{5}}
\or
  \newcommand{\miniscule}{\@setfontsize\miniscule{5}{6}}
\or
  \newcommand{\miniscule}{\@setfontsize\miniscule{5}{6}}
\fi
\makeatother

\DeclareRobustCommand{\optbar}[1]{\shortstack{{\miniscule (\rule[.5ex]{1.25em}{.18mm})}
  \\ [-.7ex] $#1$}}

\def\mumu       {{\ensuremath{\Pmu^+\Pmu^-}}\xspace}

\def\cquark    {{\ensuremath{\Pc}}\xspace}

\def\bquark    {{\ensuremath{\Pb}}\xspace}

\def\pion   {{\ensuremath{\Ppi}}\xspace}

\def\pip    {{\ensuremath{\pion^+}}\xspace}
\def\pim    {{\ensuremath{\pion^-}}\xspace}
\def\pipm   {{\ensuremath{\pion^\pm}}\xspace}

\def\kaon    {{\ensuremath{\PK}}\xspace}
  \def\Kbar    {{\kern 0.2em\overline{\kern -0.2em \PK}{}}\xspace}

\def\KorKbar    {\kern 0.18em\optbar{\kern -0.18em K}{}\xspace}

\def\Kp      {{\ensuremath{\kaon^+}}\xspace}
\def\Km      {{\ensuremath{\kaon^-}}\xspace}

\def\KS      {{\ensuremath{\kaon^0_{\mathrm{ \scriptscriptstyle S}}}}\xspace}

  \def\Dbar    {{\kern 0.2em\overline{\kern -0.2em \PD}{}}\xspace}

\def\DorDbar    {\kern 0.18em\optbar{\kern -0.18em D}{}\xspace}

\def\B       {{\ensuremath{\PB}}\xspace}
\def\Bbar    {{\ensuremath{\kern 0.18em\overline{\kern -0.18em \PB}{}}}\xspace}

\def\BorBbar    {\kern 0.18em\optbar{\kern -0.18em B}{}\xspace}

\def\Bu      {{\ensuremath{\B^+}}\xspace}

\def\jpsi     {{\ensuremath{{\PJ\mskip -3mu/\mskip -2mu\Ppsi\mskip 2mu}}}\xspace}

  \def\Y#1S{\ensuremath{\PUpsilon{(#1S)}}\xspace}

\def\proton      {{\ensuremath{\Pp}}\xspace}

\def\Lz          {{\ensuremath{\PLambda}}\xspace}
\def\Lbar        {{\ensuremath{\kern 0.1em\overline{\kern -0.1em\PLambda}}}\xspace}
\def\LorLbar    {\kern 0.18em\optbar{\kern -0.18em \PLambda}{}\xspace}

\def\Lb      {{\ensuremath{\Lz^0_\bquark}}\xspace}

\def\Lc      {{\ensuremath{\Lz^+_\cquark}}\xspace}

\newcommand{\decay}[2]{\mbox{\ensuremath{#1\!\to #2}}\xspace}         

\def\to                 {\ensuremath{\rightarrow}\xspace}

\def\AT#1     {\ensuremath{A_{\mathrm{T}}^{#1}}\xspace}           

\def\C#1      {\ensuremath{\mathcal{C}_{#1}}\xspace}                       
\def\Cp#1     {\ensuremath{\mathcal{C}_{#1}^{'}}\xspace}                    
\def\Ceff#1   {\ensuremath{\mathcal{C}_{#1}^{\mathrm{(eff)}}}\xspace}        
\def\Cpeff#1  {\ensuremath{\mathcal{C}_{#1}^{'\mathrm{(eff)}}}\xspace}       
\def\Ope#1    {\ensuremath{\mathcal{O}_{#1}}\xspace}                       
\def\Opep#1   {\ensuremath{\mathcal{O}_{#1}^{'}}\xspace}                    

\newcommand{\tev}{\ifthenelse{\boolean{inbibliography}}{\ensuremath{~T\kern -0.05em eV}}{\ensuremath{\mathrm{\,Te\kern -0.1em V}}}\xspace}
\newcommand{\Tev}{\ifthenelse{\boolean{inbibliography}}{\ensuremath{~T\kern -0.05em eV}}{\ensuremath{\mathrm{\,Te\kern -0.1em V}}}\xspace}
\newcommand{\gev}{\ensuremath{\mathrm{\,Ge\kern -0.1em V}}\xspace}
\newcommand{\mev}{\ensuremath{\mathrm{\,Me\kern -0.1em V}}\xspace}
\newcommand{\kev}{\ensuremath{\mathrm{\,ke\kern -0.1em V}}\xspace}
\newcommand{\ev}{\ensuremath{\mathrm{\,e\kern -0.1em V}}\xspace}
\newcommand{\gevc}{\ensuremath{{\mathrm{\,Ge\kern -0.1em V\!/}c}}\xspace}
\newcommand{\mevc}{\ensuremath{{\mathrm{\,Me\kern -0.1em V\!/}c}}\xspace}
\newcommand{\gevcc}{\ensuremath{{\mathrm{\,Ge\kern -0.1em V\!/}c^2}}\xspace}
\newcommand{\gevgevcccc}{\ensuremath{{\mathrm{\,Ge\kern -0.1em V^2\!/}c^4}}\xspace}
\newcommand{\mevcc}{\ensuremath{{\mathrm{\,Me\kern -0.1em V\!/}c^2}}\xspace}

\def\invfb   {\ensuremath{\mbox{\,fb}^{-1}}\xspace}

\newcommand{\chisq}{\ensuremath{\chi^2}\xspace}

\newcommand{\chisqip}{\ensuremath{\chi^2_{\text{IP}}}\xspace}

\def\gsim{{~\raise.15em\hbox{$>$}\kern-.85em
          \lower.35em\hbox{$\sim$}~}\xspace}
\def\lsim{{~\raise.15em\hbox{$<$}\kern-.85em
          \lower.35em\hbox{$\sim$}~}\xspace}

\def\ptot       {\ensuremath{p}\xspace}

\def\tell1  {TELL1\xspace}
\def\ukl1   {UKL1\xspace}

\newcommand{\eg}{\mbox{\itshape e.g.}\xspace}

\usepackage{cite} 
\usepackage{mciteplus}

\usepackage{rotating}

\begin{document}

\renewcommand{\thefootnote}{\fnsymbol{footnote}}
\setcounter{footnote}{1}


\begin{titlepage}
\pagenumbering{roman}

\vspace*{-1.5cm}
\centerline{\large EUROPEAN ORGANIZATION FOR NUCLEAR RESEARCH (CERN)}
\vspace*{0.5cm}
\noindent
\begin{tabular*}{\linewidth}{lc@{\extracolsep{\fill}}r@{\extracolsep{0pt}}}
\ifthenelse{\boolean{pdflatex}}
{\vspace*{-1.5cm}\mbox{\!\!\!\includegraphics[width=.14\textwidth]{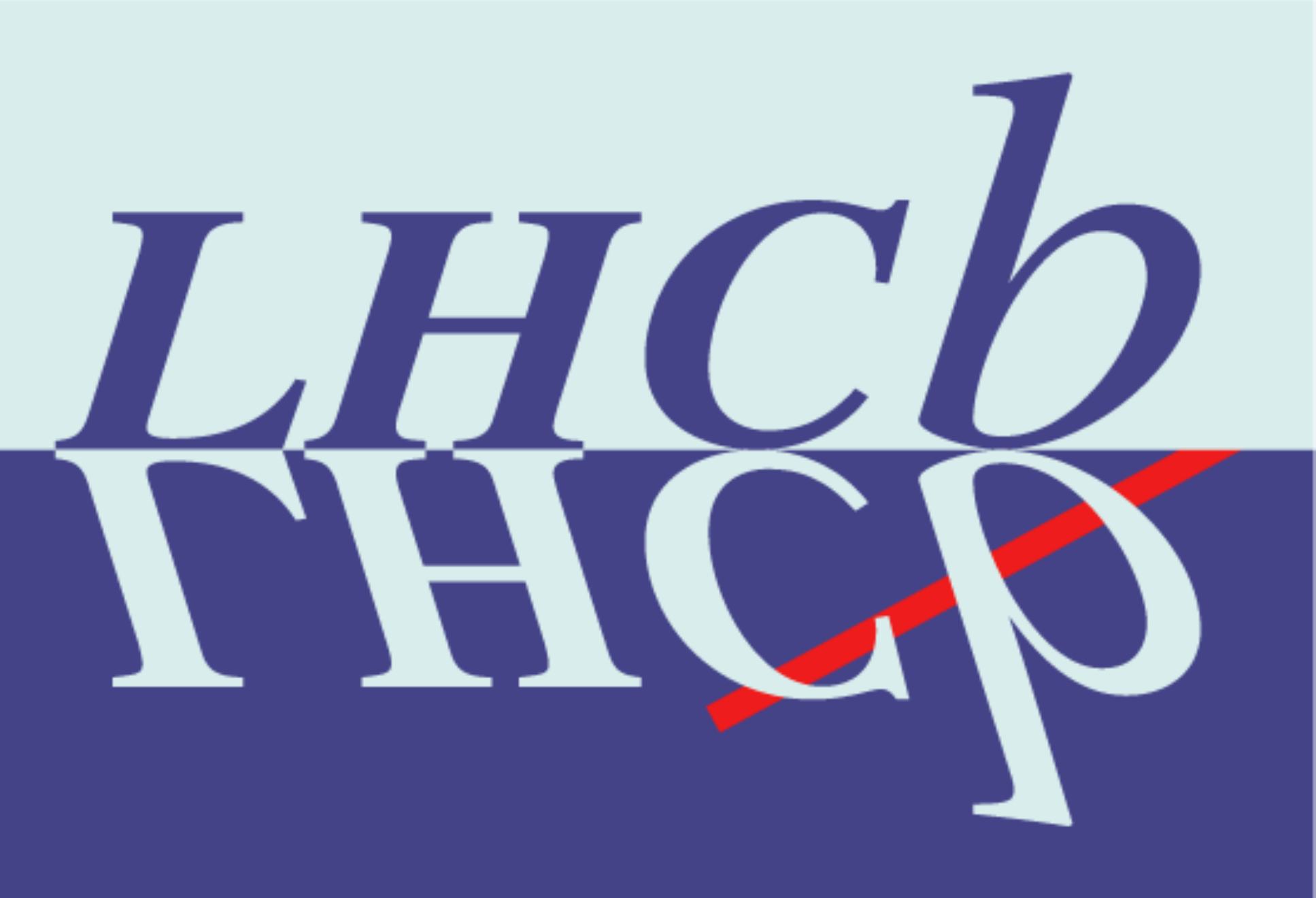}} & &}%
{\vspace*{-1.2cm}\mbox{\!\!\!\includegraphics[width=.12\textwidth]{lhcb-logo.eps}} & &}%
\\
 & & CERN-EP-2019-153 \\  
 & & LHCb-PAPER-2019-025 \\  
%
%
%
   & & July 31, 2019 \\ 
%
%
%
\end{tabular*}

\vspace*{2.0cm}

{\normalfont\bfseries\boldmath\huge
\begin{center}
  \papertitle 
\end{center}
}

\vspace*{1.0cm}

\begin{center}
\paperauthors\footnote{Authors are listed at the end of this Letter.}
\end{center}

\vspace{\fill}

\begin{abstract}
\noindent
We report the observation of a new structure in the $\Lb \pip\pim$~spectrum using 
the~full LHCb data set
of $\proton\proton$~collisions, 
corresponding to an integrated luminosity 
of~9\invfb, collected at 
$\sqrt{s}=7$, 8 and $13\,\mathrm{TeV}$.
A~study of the~structure suggests its interpretation as a superposition of two almost degenerate narrow states.
The~masses and widths of these states are measured to be
\begin{eqnarray*}
    m_{\Lambda_{\bquark}\mathrm{(6146)}^0}    &=&6146.17 \pm 0.33 \pm 0.22 \pm 0.16 \mev \,, \\
    m_{\Lambda_{\bquark}\mathrm{(6152)}^0}    &=&6152.51 \pm 0.26 \pm 0.22 \pm 0.16 \mev \,, \\
\Gamma_{\Lambda_{\bquark}\mathrm{(6146)}^0}&=&\phantom{615}2.9\phantom{0}\pm 1.3\phantom{0} \pm 0.3\phantom{0}\mev\,, \\
\Gamma_{\Lambda_{\bquark}\mathrm{(6152)}^0}      &=&\phantom{615}2.1\phantom{0}\pm 0.8\phantom{0} \pm 0.3\phantom{0}\mev \,, 
\end{eqnarray*}
with a~mass~splitting  of $\Delta m = 6.34 \pm 0.32\pm0.02\mev$,
where the first uncertainty is statistical, the second systematic 
and the third derives from the knowledge of the mass of the \Lb~baryon.
The~measured masses and widths of these new excited states suggest their possible interpretation as a~doublet of $\Lambda_{\bquark}\mathrm{(1D)}^0$~states. 

\end{abstract}

\vspace*{1.0cm}

\begin{center}
  Published in \href{https://doi.org/10.1103/PhysRevLett.123.152001}{Phys.\ Rev.\ Lett.~{\bf{123}}, 152001\,(2019).}
\end{center}

\vspace{\fill}

{\footnotesize 
\centerline{\copyright~\papercopyright, \href{\paperlicenceurl}{\paperlicence}.}}
\vspace*{2mm}

\end{titlepage}


\newpage
\setcounter{page}{2}
\mbox{~}
%

%
%
%

\cleardoublepage


\renewcommand{\thefootnote}{\arabic{footnote}}
\setcounter{footnote}{0}



\pagestyle{plain} 
\setcounter{page}{1}
\pagenumbering{arabic}


%

In the constituent quark model~\cite{GELLMANN1964214,Zweig:352337}, baryons containing 
a~beauty quark form multiplets according to the~internal symmetries of flavour, 
spin, and parity~\cite{RevModPhys.82.1095}. Beyond the~\Lb baryon, which is the~lightest beauty baryon, 
a~rich spectrum of radially and orbitally excited states is expected at higher masses. 
Several new baryon states have been discovered in recent 
years~\cite{LHCb-PAPER-2012-012,LHCb-PAPER-2018-013,PhysRevLett.108.252002,LHCb-PAPER-2014-061,LHCb-PAPER-2018-032}.
The spectrum of excited states decaying to the $\Lb\pip\pim$ final 
state has already been studied by the \lhcb experiment with 
the~discovery 
of two narrow states~\cite{LHCb-PAPER-2012-012}, 
denoted $\Lambda_{\bquark}\mathrm{(5912)}^0$ 
and $\Lambda_{\bquark}\mathrm{(5920)}^0$.
The heavier of these states was later confirmed by the~\cdf collaboration~\cite{Aaltonen:2013tta}. 
Mass predictions for the ground-state beauty baryons 
and their orbital and radial excitations 
are given in many theoretical works, \eg,~\cite{Chen:2014nyo, Ebert:2011kk, Roberts:2007ni, Capstick:1986bm}. 
In~addition to the already observed doublet of first orbital excitations, 
more states are predicted in the~mass region near or above \ifthenelse{\boolean{prl}}{6.1\gev\,(natural  
units with $c=\hbar=1$ are used throughout this Letter).}{
6.1\gev.\footnote{Natural  
units with $c=\hbar=1$ are used throughout this Letter.}}

In this Letter, we document the study of the $\Lb\pip\pim$ 
spectrum\,(charge conjugation is implied throughout this article) 
in the~extended mass region 
between 6.10 and~6.25\gev, 
using $\proton\proton$~collision data collected 
by 
the~\lhcb experiment at centre\nobreakdash-of\nobreakdash-mass energies of  7, 8, and 13\,$\mathrm{TeV}$. 
The~combined data set  corresponds to an integrated luminosity of 9\invfb.

The \lhcb detector~\cite{Alves:2008zz,LHCb-DP-2014-002} is a~single\nobreakdash-arm forward
spectrometer covering the~\mbox{pseudorapidity} range~\mbox{$2<\eta <5$},
designed for the study of particles containing \bquark~or \cquark~quarks.
The detector includes a~high\nobreakdash-precision tracking system
consisting of a~silicon\nobreakdash-strip vertex detector surrounding 
the~$\proton\proton$~interaction region~\cite{LHCb-DP-2014-001}, 
a~large\nobreakdash-area silicon\nobreakdash-strip detector located
upstream of a dipole magnet with a bending power of about
$4{\mathrm{\,Tm}}$, and three stations of silicon-strip detectors and straw
drift tubes~\cite{LHCb-DP-2013-003} placed downstream of the~magnet.
The tracking system provides a measurement of the~momentum, \ptot, of charged particles with
a relative uncertainty that varies from 0.5\% at low momentum to 1.0\% at 200\gev.
The~momentum scale of the~tracking system is calibrated using samples of 
$\decay{\jpsi}{\mumu}$
and $\decay{\Bu}{\jpsi\Kp}$ decays collected concurrently with 
the~data sample
used for this analysis~\cite{LHCb-PAPER-2012-048,LHCb-PAPER-2013-011}.
The~relative accuracy of this procedure is estimated to be
$3 \times 10^{-4}$ using samples of other fully reconstructed
$\bquark$-hadron, $\KS$, and narrow \mbox{$\PUpsilon\mathrm{(1S)}$}~resonance decays.
Different types of charged hadrons are distinguished using information
from two ring-imaging Cherenkov detectors~\cite{LHCb-DP-2012-003}.
The~online event selection is performed by a~trigger~\cite{LHCb-DP-2012-004}
which consists of a~hardware stage, based on information from the calorimeter and muon systems,
followed by a software stage, which applies a full event reconstruction.
The software trigger requires a two-, three- or four-track secondary vertex with significant displacement from all primary $\proton\proton$~interaction vertices. 
A~multivariate algorithm~\cite{BBDT} 
is used for the identification of secondary vertices consistent with the decay 
of a~$\bquark$~hadron.
Simulated data samples are produced using the software packages described in
Refs.~\cite{Sjostrand:2006za,*Sjostrand:2007gs,LHCb-PROC-2010-056,Lange:2001uf,Golonka:2005pn,Allison:2006ve,*Agostinelli:2002hh}.

Samples of \Lb~candidates are formed from $\Lc\pim$ combinations,
where the \Lc~baryon is reconstructed in the~$\proton\Km\pip$~final state.
All charged final\nobreakdash-state particles are required to
have particle-identification information consistent with their respective 
mass hypotheses.
Misreconstructed tracks are suppressed by the use of a~neural network~\cite{DeCian:2255039}.
To~suppress prompt background, the~\Lb~decay products are required to 
have significant \chisqip with respect to all PVs in the~event,
where \chisqip of a~particle is the~difference 
in \chisq of the~vertex fit of a~given PV, 
when the~particle is included or excluded from the~fit.
The~reconstructed \Lc vertex is required to have a good fit quality and to be significantly
displaced from all PVs. 
The~reconstructed \Lc~mass must be within
a~mass window of $\pm 25$\mev of the~known value~\cite{PDG2018}.
Pion candidates 
are combined with \Lc candidates to form \Lb~candidates, 
requiring good vertex-fit quality
and separation of the~\Lb decay point from any PV in the~event.
A~Boosted Decision Tree\,(BDT) discriminant~\cite{ROE2005577,AdaBoost} 
is used to further reduce the~background level.
The~BDT exploits fifteen variables, 
including kinematic variables of 
the~$\Lc$ 
and $\Lb$~candidates, 
the~lifetime of the~$\Lb$~candidate, 
kinematic variables and quality of 
particle identification for the~final\nobreakdash-state pions, kaons and protons, 
and variables describing 
the~consistency of the~selected candidates with the~$\decay{\Lb}{\Lc\pim}$~decay 
of a~\Lb~baryon~\cite{Hulsbergen:2005pu}.
The~BDT is trained using background\nobreakdash-subtracted~\cite{Pivk:2004ty}
\Lb~candidates as a~signal sample
and \Lb~candidates from the~data sidebands, in the~$\Lc\pim$~mass range \mbox{$5.7< m_{\Lc\pim}<6.1\gev$}, as a~background sample.
The~$k$\nobreakdash-fold cross\nobreakdash-validation
technique with $k=11$ is used in the~BDT 
training~\cite{geisser1993predictive}.
The~use of a~multivariate discriminant allows the~mall level of \Lb~background candidates in the~analysis to be reduced by a~further factor of two, 
keeping almost 100\% efficiency for the~signal.
The~resulting yield of $\decay{\Lb}{\Lc\pim}$~decays is $(892.8\pm1.2)\times10^3$. 
A~sample of $\decay{\Lb}{\jpsi\proton\Km}$~candidates, 
with $\decay{\jpsi}{\mumu}$,  
is also selected in a similar way as a~cross\nobreakdash-check. 
The~yield for this decay mode is smaller,  
corresponding to \mbox{$(217.5\pm0.7)\times10^3$} decays..  
The~mass spectra of the~selected $\decay{\Lb}{\Lc\pim}$ and $\decay{\Lb}{\jpsi\proton\Km}$~candidates are 
shown in 
Fig.~\ref{fig:Lb}
of the~Supplemental Material of this Letter.

The~selected $\Lb$~candidates are combined with 
pairs of pions compatible with originating from the~same PV
as the~\Lb~candidate.
Only pion pairs with \mbox{$p_{\mathrm{T}}^{\pip\pim}>500\mev$} are used, to suppress
the~otherwise large combinatorial background from soft dipion combinations.
This background is further reduced by using a~dedicated BDT discriminant 
tuned on each of the~two samples 
with $\decay{\Lb}{\Lc\pim}$ and $\decay{\Lb}{\jpsi\proton\Km}$~decays.
It~exploits the~transverse momentum of the~$\Lb\pip\pim$~combination, 
the~$\chisq$~value for the~$\Lb\pip\pim$~vertex, 
the~transverse momenta of both individual pions and the~pion pair, 
as well as particle\nobreakdash-identification and 
reconstruction\nobreakdash-quality~\cite{DeCian:2255039}
variables for both pions. The~BDT is trained 
on simulated samples of excited beauty baryons with a~mass of 6.15\gev as signal
and \emph{same\nobreakdash-sign} $\Lb\Ppi^{\pm}\Ppi^{\pm}$~combinations in 
data, with \mbox{$m_{\Lb\Ppi^{\pm}\Ppi^{\pm}}<6.22\gev$}, as background.
In~simulation unpolarized production
of excited beauty baryons is assumed,
followed by decays to the~$\Lb\pip\pim$~final state according
to a~three\nobreakdash-body phase\nobreakdash-space decay model.

In~order to improve the~$\Lb\pip\pim$~mass resolution, 
the~$\Lb\pip\pim$ combinations are refitted constraining the~masses of 
the~\Lc~baryon\,(or \jpsi~meson) to their known values~\cite{PDG2018}
and requiring consistency of the~$\Lb\pip\pim$ vertex 
with the~PV 
associated with the~\Lb~candidate~\cite{Hulsbergen:2005pu}.
The~mass of the~\Lb~baryon in the~fit is 
constrained  to the~central value of 
\mbox{$m_{\Lb}=5618.62 \pm 0.16 \pm 0.13\mev$}~\cite{LHCb-PAPER-2017-011}, 
 obtained from a~combination of the~measurements of
 the~\Lb~mass  in 
$\decay{\Lb}{\Pchi_{\cquark1,2}\proton\Km}$~\cite{LHCb-PAPER-2017-011}, 
$\decay{\Lb}{\Ppsi\mathrm{(2S)}\proton\Km}$,
$\decay{\Lb}{\jpsi\pip\pim\proton\Km}$~\cite{LHCb-PAPER-2015-060}
and $\decay{\Lb}{\jpsi\Lambda}$~decay
modes~\cite{LHCb-PAPER-2011-035,LHCb-PAPER-2012-048} by the~LHCb collaboration.
The~mass distributions for selected $\Lb\pip\pim$~candidates
are shown in Fig.~\ref{fig:figure1}.
Only $\Lb$~candidates with a~mass within 
$\pm50\,(20)\mev$\,(approximately three times the~resolution) 
of the known \Lb~mass 
for $\decay{\Lb}{\Lc\pim}\,(\decay{\Lb}{\jpsi\proton\Km})$~candidates
are used.
There is a~clear excess of $\Lb\pip\pim$ candidates around $6.15\gev$
over the background for both \Lb~decay modes. 
The~excess is initially treated as originating 
from a~single broad state.
The~distributions are parameterised by the~sum of signal and background components.
The~signal component is modelled  by a~relativistic 
S\nobreakdash-wave Breit$-$Wigner function
with Blatt$-$Weisskopf form factors~\cite{Blatt:1952ije}. 
The~relativistic Breit$-$Wigner function
is convolved with the~detector resolution described by the~sum of two Gaussian functions
with common mean and parameters, which are fixed from simulation.
The~obtained effective resolution is 1.7\mev. 
The~background component is parameterised with a~second\nobreakdash-order polynomial function.
Extended unbinned maximum\nobreakdash-likelihood  fits to the~$\Lb\pip\pim$~mass spectra are shown in~Fig.~\ref{fig:figure1}.
The~corresponding parameters of interest are listed in Table~\ref{tab:table1}.


\ifthenelse{\boolean{prl}}{
\begin{figure}[tbh]
  \centering
      \includegraphics*[width=\linewidth,
       ]{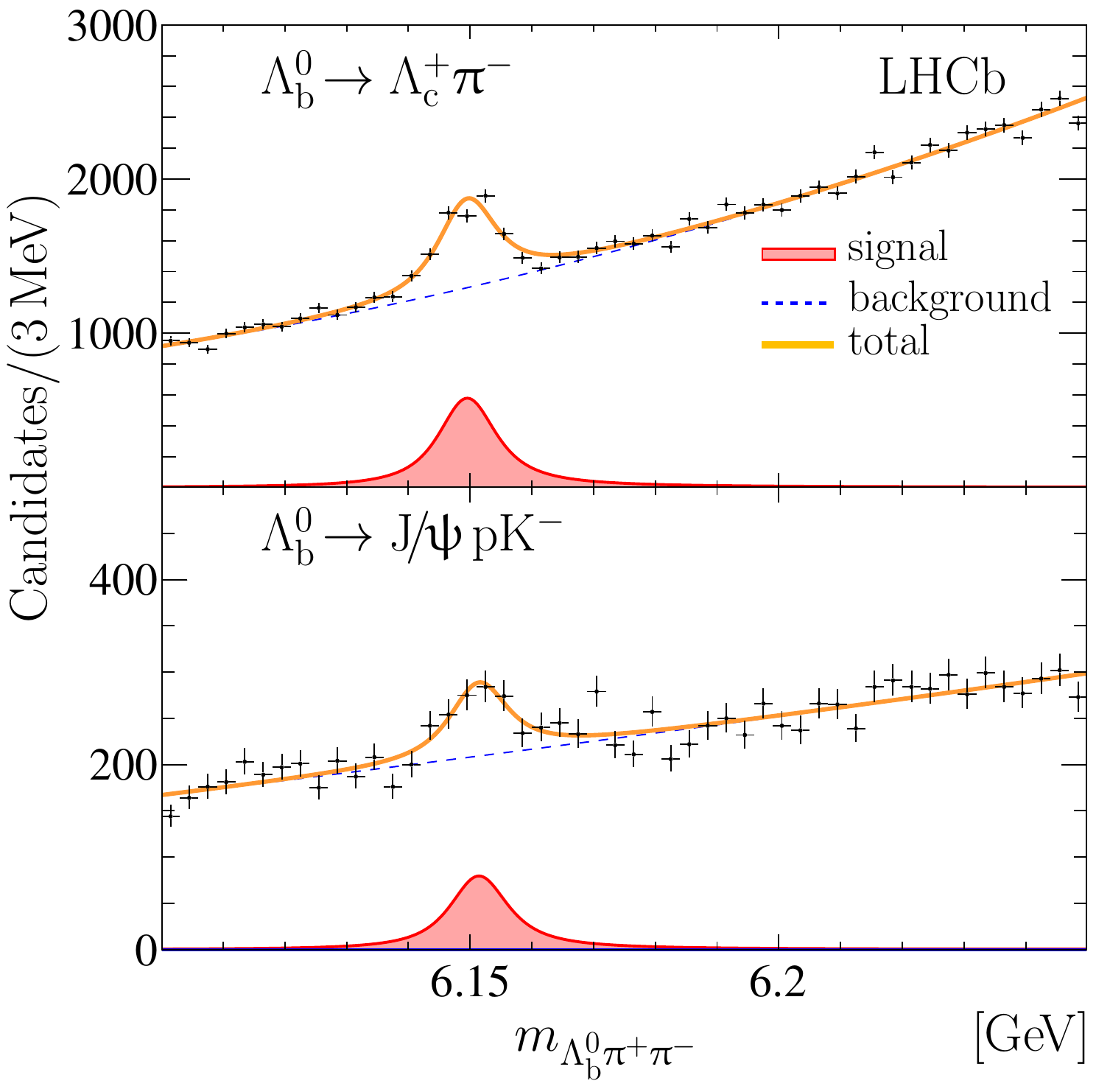}
  \caption { \small
  Mass distribution of selected $\Lb\pip\pim$~candidates
  for the (top) $\decay{\Lb}{\Lc\pim}$
  and (bottom) $\decay{\Lb}{\jpsi\proton\Km}$ decay modes.
  }
  \label{fig:figure1}
\end{figure}
}{
\begin{figure}[t]
  \setlength{\unitlength}{1mm}
  \centering
  \begin{picture}(150,150)
    %
    \put(  0, 0){ 
      \includegraphics*[width=150mm,height=150mm,%
       ]{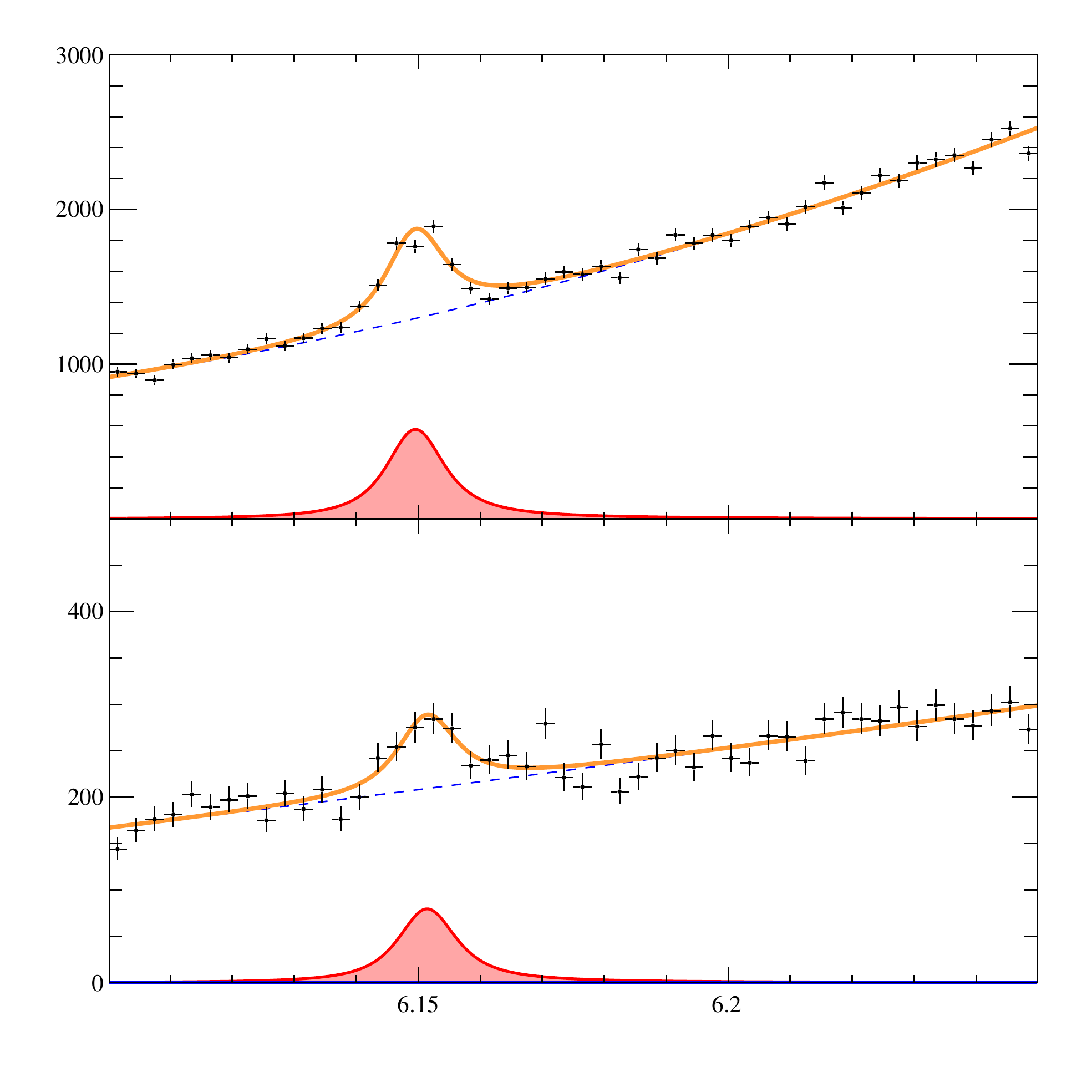}
    }
    \put(109,110) {\begin{tikzpicture}[x=1mm,y=1mm]\filldraw[fill=red!35!white,draw=red,thick]  (0,0) rectangle (10,1.5);\end{tikzpicture} }
    \put(109,105) {\color[rgb]{0.00,0.00,1.00} {\hdashrule[0.5ex][x]{10mm}{0.5mm}{1.0mm 1.0mm} } } 
    \put(109,100) {\color[rgb]{1.00,0.75,0.00} {\hdashrule[0.5ex][x]{10mm}{1.0mm}{1.0mm 0.0mm} } } 
    \put(120,110) {signal}
    \put(120,105) {background}
    \put(120,100) {total}

    \put(125,135){\lhcb}
   \put(2,60){\begin{sideways}Candidates/(3\mev)\end{sideways}}
    \put(30,135){$\decay{\Lb}{\Lc\pim}$}
    \put(30, 72){$\decay{\Lb}{\jpsi\proton\Km}$}
  \put( 70,  3) {$m_{\Lb\pip\pim}$} \put( 134,  3) {$\left[\!\gev\right]$}
  \end{picture}
  \caption { \small
    Mass distribution of selected $\Lb\pip\pim$~candidates
  for the (top) $\decay{\Lb}{\Lc\pim}$
  and (bottom) $\decay{\Lb}{\jpsi\proton\Km}$ decay modes.
  }
  \label{fig:figure1}
\end{figure}
}


\ifthenelse{\boolean{prl}}{
\begin{table}[htb]
  \centering
  \caption{\small
   The  yields, $N$, masses, $m$,  and natural 
  widths, $\Gamma$, 
  from the~fits of a~single broad state 
  to the~$\Lb\pip\pim$~mass spectra.
  }\label{tab:table1}
  \vspace*{0.5mm}
  \begin{tabular*}{0.9\linewidth}{@{\hspace{2mm}}l@{\extracolsep{\fill}}cc@{\hspace{2mm}}}
    & $\decay{\Lb}{\Lc\pim}$
    & $\decay{\Lb}{\jpsi\proton\Km}$
    \\[1mm]
    \hline 
    \\[-2mm]
   $N_{\Lb\pip\pim}$  
   & $\phantom{0.0}3117\pm240\phantom{.}$ & $\phantom{0.00}431\pm97\phantom{.0}$ 
   \\
   $m~\left[\mev\right]$\
   & $6149.64\pm0.34\phantom{}$   & $6151.51\pm0.97\phantom{}$ 
   \\
  $\Gamma~\left[\mev\right]$
   & $\phantom{000}9.61\pm0.98\phantom{}$ & $\phantom{000}9.67\pm2.89\phantom{}$
   \end{tabular*}
\end{table}
}{
\begin{table}[b]
  \centering
  \caption{\small
  The  yields, $N$, masses, $m$, and natural 
  widths, $\Gamma$, 
  from the~fits of a~single broad state 
  to the~$\Lb\pip\pim$~mass spectra.
  }\label{tab:table1}
  \vspace*{0.5mm}
  \begin{tabular*}{0.75\linewidth}{@{\hspace{3mm}}l@{\extracolsep{\fill}}lcc@{\hspace{3mm}}}
    & 
    & $\decay{\Lb}{\Lc\pim}$ mode
    & $\decay{\Lb}{\jpsi\proton\Km}$ mode
    \\
    \hline 
   $N_{\Lb\pip\pim}$ &   
   & $\phantom{0.0}3117\pm240\phantom{.}$ & $\phantom{0.00}431\pm97\phantom{0.}$ 
   \\
   $m$ & $\left[\mev\right]$\
   & $6149.6\pm0.3\phantom{}$  
   & $6151.5\pm1.0\phantom{}$ 
   \\
  $\Gamma$ & $\left[\mev\right]$
     & $\phantom{000}9.6\pm1.0\phantom{}$ & $\phantom{000}9.7\pm2.9\phantom{}$
   \end{tabular*}
\end{table}
}

The~mass and width of the~structure agree between 
the~$\decay{\Lb}{\Lc\pim}$ and $\decay{\Lb}{\jpsi\proton\Km}$~samples.
The~statistical significance for the~signals
is estimated using Wilks' theorem~\cite{Wilks:1938dza}. 
It~is found to exceed twenty\nobreakdash-six and nine standard deviations
for the~$\decay{\Lb}{\Lc\pim}$ 
and $\decay{\Lb}{\jpsi\proton\Km}$~decay modes, respectively. 
The~fitted parameters exhibit very modest dependence 
on the~choice of the~orbital momentum 
for the~relativistic Breit$-$Wigner function and 
the~Blatt$-$Weiskopf  breakup momenta~\cite{Blatt:1952ije}.
The~signal yields, masses and widths are found to be
consistent for~the different data\nobreakdash-taking periods
and between the~$\Lb\pip\pim$ and 
$\bar{\Lambda}_{\bquark}^0\pip\pim$~final states.

Since the~mass of the~new structure is above 
the~$\Sigma_{\bquark}^{(\ast)\pm}\Ppi^{\mp}$~kinematic thresholds, 
the~$\Lb\pip\pim$~mass spectrum is investigated 
in $\Lb\Ppi^{\pm}$~mass regions
populated by the~$\Sigma_{\bquark}^{(\ast)\pm}$~resonances.
The~data are split into three 
nonoverlapping regions:
candidates with a~$\Lb\Ppi^{\pm}$ mass 
within the~natural width
of the~known $\Sigma_{\bquark}^{\pm}$~mass;
candidates with a~$\Lb\Ppi^{\pm}$ mass 
within the~natural width 
of the~known $\Sigma_{\bquark}^{\ast\pm}$~mass;
and the~remaining nonresonant\,(NR) region.
The~$\Lb\pip\pim$~mass spectra in these three regions 
are shown in Fig.~\ref{fig:figure2}. 
Only~the~larger sample of $\Lb$~candidates
selected via the $\decay{\Lb}{\Lc\pim}$~decay mode is used
here and in the~remainder of this Letter.
The~spectra in 
the~$\Sigma_{\bquark}$ and $\Sigma_{\bquark}^{\ast}$~regions
look different and suggest 
the~presence of two narrow peaks.

Doublets of orbitally excited states are predicted in the mass region near 
the~observed peaks~\cite{Chen:2014nyo, Ebert:2011kk, Roberts:2007ni, Capstick:1986bm}. 
The~spins and parities of the states in the~doublet determine 
the~lowest allowed orbital angular momentum in the~two\nobreakdash-body 
$\Sigma_{\bquark}^{(\ast)\pm}\Ppi^{\mp}$~transition.
The~intensities of the~transitions  can be enhanced or 
suppressed depending on the~angular momentum assignment.
Heavy quark effective theory\,(HQET) 
also predicts different decay rates of the~doublet members to 
the~$\Sigma_{\bquark}^{\pm}\Ppi^{\mp}$ and $\Sigma^{\ast\pm}_{\bquark}\Ppi^{\mp}$~final states~\cite{Isgur:1991wq}.
To~probe the~two\nobreakdash-resonance hypothesis, 
a~simultaneous fit to the~mass spectra 
in the~three $\Lb\pipm$ mass regions is performed.
For~each region, the~fit function consists of 
two signal components and a~background component described 
by a~second\nobreakdash-order polynomial function. 
The~signal components are modelled by relativistic Breit$-$Wigner
functions convolved with the~detector resolution.
For~the~$\Sigma_{\bquark}$~region, 
the~signal components describe  
two\nobreakdash-body intermediate states 
$\Sigma_{\bquark}^{\pm}\Ppi^{\mp}$
in P\nobreakdash- and D\nobreakdash-wave for the~low\nobreakdash-mass and high\nobreakdash-mass states, respectively.
For~the~$\Sigma_{\bquark}^{\ast}$~region, 
S\nobreakdash- and P\nobreakdash-wave are chosen for decays of low\nobreakdash- and high\nobreakdash-mass states, respectively.
These choices are motivated by the~possible 
interpretation of the new states as a~doublet of
$\Lambda_{\bquark}\mathrm{(1D)^0}$~states~\mbox{\cite{Chen:2014nyo, Ebert:2011kk, Roberts:2007ni, Capstick:1986bm}}. 
The~masses and widths of the~two states are taken as 
common parameters
for the~three regions, while the~other parameters, 
namely the~signal and background yields and 
background shape parameters, are allowed to vary independently.
The~two signal components are added incoherently, 
\mbox{assuming} interference effects are negligible, 
since a~coherent production of the~states in 
the~complex environment of $\proton\proton$~interactions 
is unlikely.

\ifthenelse{\boolean{prl}}{
\begin{figure}[t]
  \centering
      \includegraphics*[width=\linewidth,
       ]{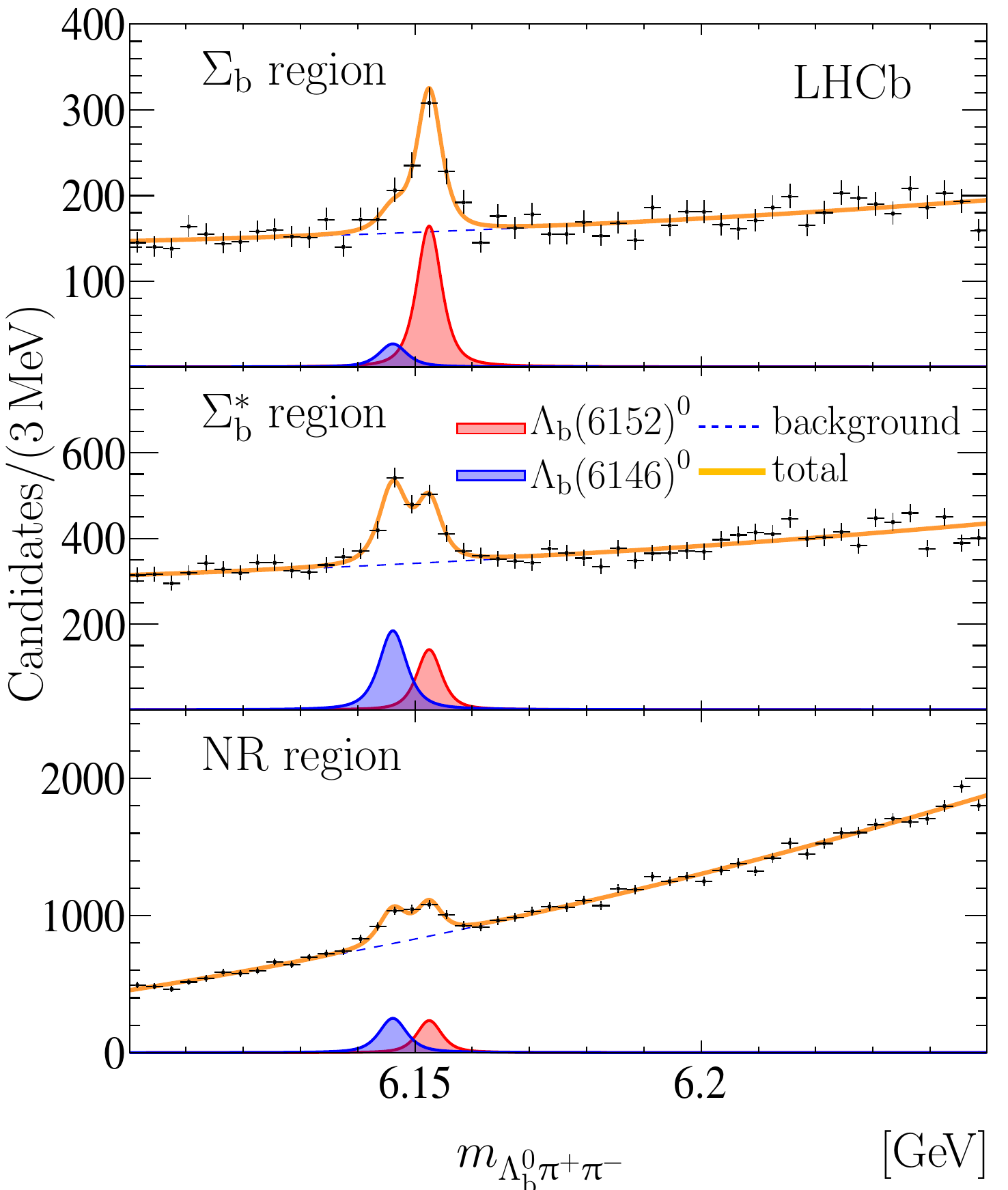}
  \caption { \small
  Mass distributions of selected $\Lb\pip\pim$~candidates
  for the three regions in $\Lb\Ppi^{\pm}$~mass:
  (top)~$\Sigma_{\bquark}$,
  (middle)~$\Sigma_{\bquark}^{\ast}$
  and (bottom)~nonresonant\,(NR) region.
  }
   \label{fig:figure2}
\end{figure}
}{
\begin{figure}[t]
  \setlength{\unitlength}{1mm}
  \centering
  \begin{picture}(150,180)
    %
    \put(  0, 0){ 
      \includegraphics*[width=150mm,height=180mm,%
       ]{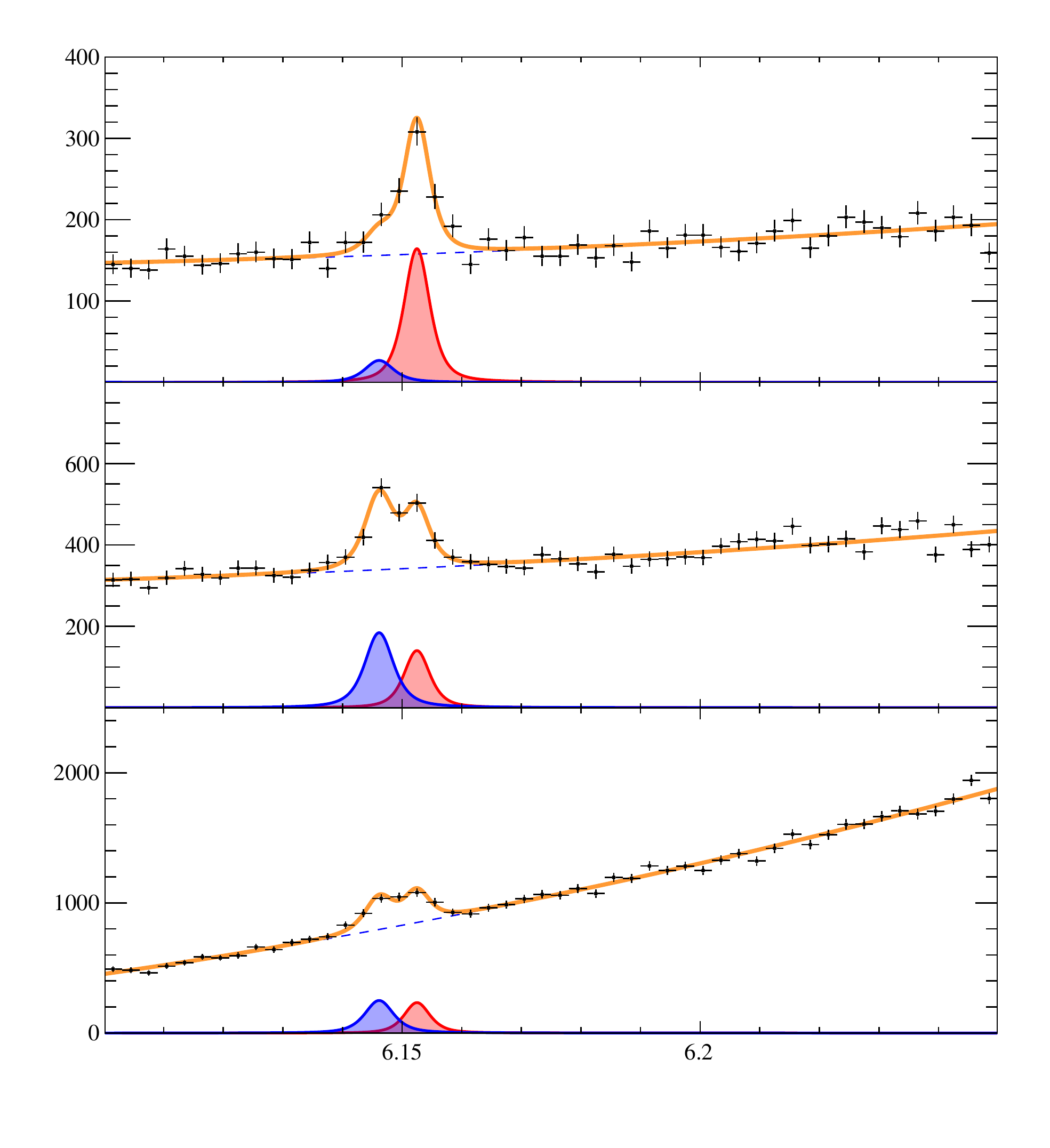}
    }
    \put(77,110) {\begin{tikzpicture}[x=1mm,y=1mm]\filldraw[fill=red!35!white,draw=red,thick]  (0,0) rectangle (10,1.5);\end{tikzpicture} }
    \put(77,105) {\begin{tikzpicture}[x=1mm,y=1mm]\filldraw[fill=blue!35!white,draw=blue,thick]  (0,0) rectangle (10,1.5);\end{tikzpicture} }
    \put(107,110) {\color[rgb]{0.00,0.00,1.00} {\hdashrule[0.5ex][x]{10mm}{0.5mm}{1.0mm 1.0mm} } } 
    \put(107,105) {\color[rgb]{1.00,0.75,0.00} {\hdashrule[0.5ex][x]{10mm}{1.0mm}{1.0mm 0.0mm} } } 
    \put( 88,110) {$\Lambda_{\bquark}\mathrm{(6152)}^0$}
    \put( 88,105) {$\Lambda_{\bquark}\mathrm{(6146)}^0$}
    \put(118,110) {$\mathrm{background}$}
    \put(118,105) {{total}}
    \put(125,163){\lhcb}
    \put(30,163){$\Sigma_{\bquark}$~region}
    \put(30,113){$\Sigma_{\bquark}^{\ast}$~region}
    \put(30, 62){NR~region}
   \put(1,70){\begin{sideways}Candidates/(3\mev)\end{sideways}}
   \put( 70,  4) {$m_{\Lb\pip\pim}$} \put( 134,  4) {$\left[\!\gev\right]$}
  \end{picture}
  \caption { \small
  Mass distributions of selected $\Lb\pip\pim$~candidates
  for the three regions in $\Lb\Ppi^{\pm}$~mass:
  (top)~$\Sigma_{\bquark}$,
  (middle)~$\Sigma_{\bquark}^{\ast}$
  and (bottom)~nonresonant\,(NR) region.
  }
  \label{fig:figure2}
\end{figure}
}


The results of the simultaneous extended unbinned maximum-likelihood 
fit to  the~$\Lb\pip\pim$~mass spectra in 
the~three $\Lb\Ppi^{\pm}$~mass regions are shown in
Fig.~\ref{fig:figure2}. 
The~two\nobreakdash-signal hypothesis is favoured with respect to 
the~single\nobreakdash-signal hypothesis 
with a~statistical significance exceeding seven~standard deviations. 
The~masses, $m$,  and the~natural widths, $\Gamma$, of 
the~two narrow states, 
referred to  hereafter 
as $\Lambda_{\bquark}\mathrm{(6146)}^0$ and 
$\Lambda_{\bquark}\mathrm{(6152)}^0$, are measured to be 
\begin{eqnarray*}
     m_{\Lambda_{\bquark}\mathrm{(6146)}^0}      & = &  6146.17 \pm 0.33 \mev \,, \\
     m_{\Lambda_{\bquark}\mathrm{(6152)}^0}      & = &  6152.51 \pm 0.26 \mev \,, \\
\Gamma_{\Lambda_{\bquark}\mathrm{(6146)}^0} 
& = &  \phantom{000}2.9\phantom{0} \pm1.3\phantom{0}\mev \,, \\ 
\Gamma_{\Lambda_{\bquark}\mathrm{(6152)}^0}       
& = &  \phantom{000}2.1\phantom{0} \pm 0.8\phantom{0}\mev \,, 
\end{eqnarray*} 
with a~mass~splitting of $\Delta m = 6.34 \pm 0.32\mev$,
where the~uncertainties are statistical only.
While these new states are denoted as $\Lambda_{\bquark}$,
their interpretation as other excited beauty baryons, such as neutral 
$\Sigma_{\bquark}^0$ states, cannot be excluded. 

\ifthenelse{\boolean{prl}}
{
\begin{figure}[tb]
  \centering
   \includegraphics*[width=\linewidth,
       ]{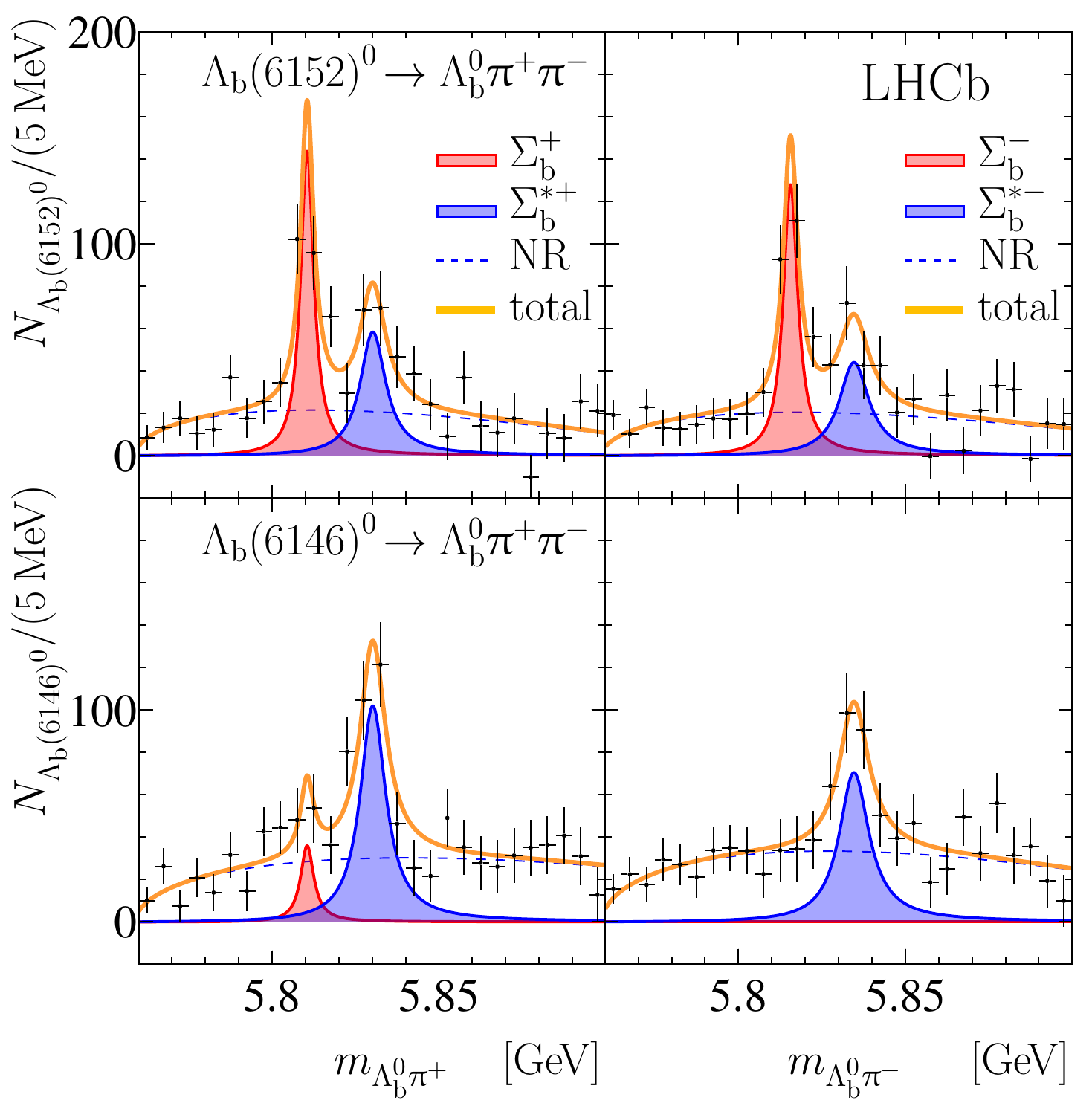}
  \caption { \small
  Background-subtracted mass distribution of (left)~$\Lb\pip$
  and (right)~$\Lb\pim$ combinations from 
  (top)~\mbox{$\decay{\Lambda_{\bquark}\mathrm{(6152)}^0}{\Lb\pip\pim}$}
  and (bottom)~\mbox{$\decay{\Lambda_{\bquark}\mathrm{(6146)}^0}{\Lb\pip\pim}$~decays}.
  Results of fits with a~model comprising  $\Sigma_{\bquark}$,
  $\Sigma_{\bquark}^{\ast}$ and nonresonant\,(NR) components 
  are superimposed.
  }
  \label{fig:figure3}
\end{figure}
}{
\begin{figure}[tb]
  \setlength{\unitlength}{1mm}
  \centering
  \begin{picture}(150,150)
    %
    \put(  0, 0){ 
      \includegraphics*[width=150mm,height=150mm,%
       ]{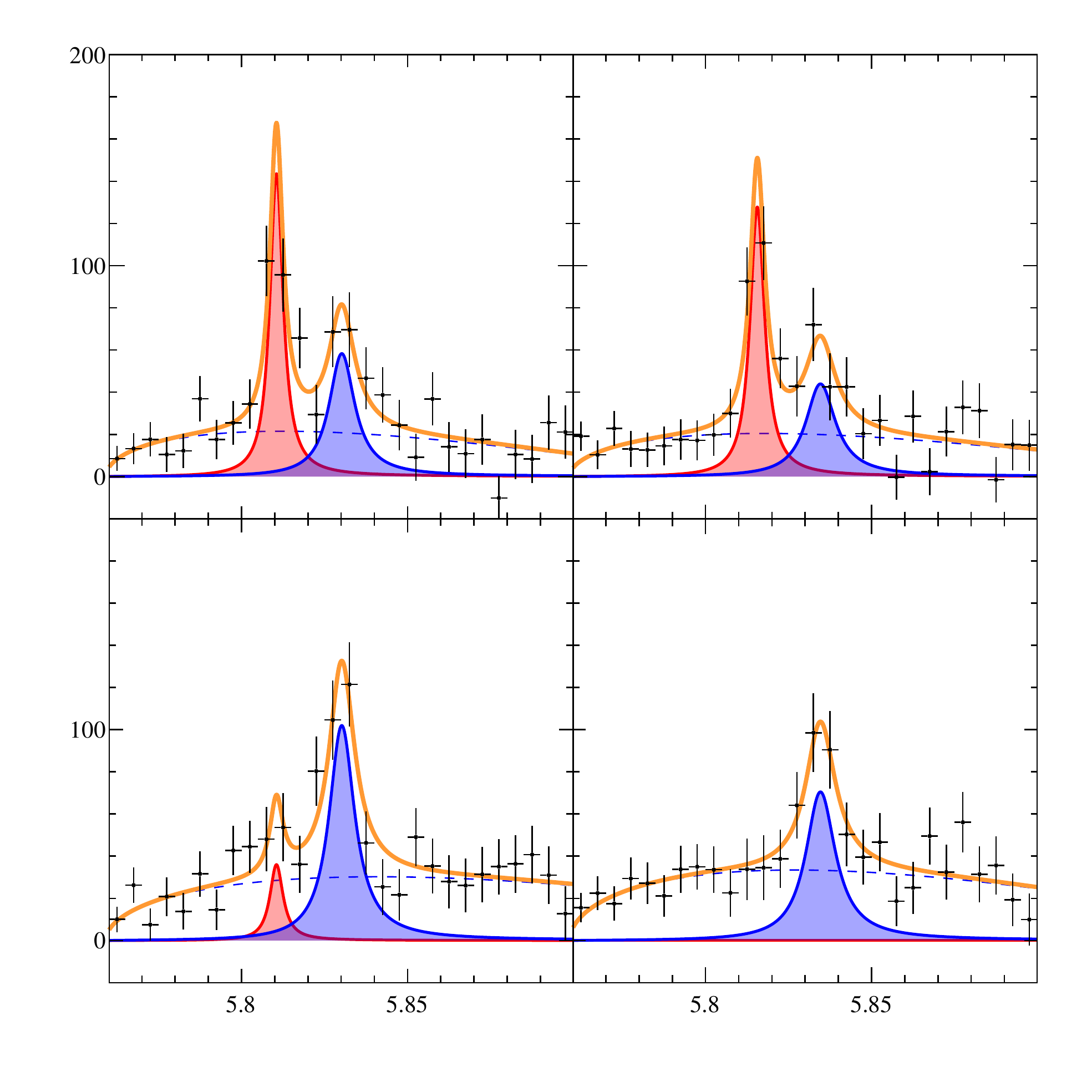}
    }
    \put(59,120) {\begin{tikzpicture}[x=1mm,y=1mm]\filldraw[fill=red!35!white,draw=red,thick]  (0,0) rectangle (8,1.5);\end{tikzpicture} }
    \put(59,115) {\begin{tikzpicture}[x=1mm,y=1mm]\filldraw[fill=blue!35!white,draw=blue,thick]  (0,0) rectangle (8,1.5);\end{tikzpicture} }
    \put(59,110) {\color[rgb]{0.00,0.00,1.00} {\hdashrule[0.5ex][x]{8mm}{0.5mm}{1.0mm 1.0mm} } } 
    \put(59,105) {\color[rgb]{1.00,0.75,0.00} {\hdashrule[0.5ex][x]{8mm}{1.0mm}{1.0mm 0.0mm} } } 
    \put( 69,120) {$\Sigma_{\bquark}^{+}$}
    \put( 69,115) {$\Sigma_{\bquark}^{\ast+}$}
    \put( 69,110) {$\mathrm{NR}$}
    \put( 69,105) {total}
    \put(123,120) {\begin{tikzpicture}[x=1mm,y=1mm]\filldraw[fill=red!35!white,draw=red,thick]  (0,0) rectangle (8,1.5);\end{tikzpicture} }
    \put(123,115) {\begin{tikzpicture}[x=1mm,y=1mm]\filldraw[fill=blue!35!white,draw=blue,thick]  (0,0) rectangle (8,1.5);\end{tikzpicture} }
    \put(123,110) {\color[rgb]{0.00,0.00,1.00} {\hdashrule[0.5ex][x]{8mm}{0.5mm}{1.0mm 1.0mm} } } 
    \put(123,105) {\color[rgb]{1.00,0.75,0.00} {\hdashrule[0.5ex][x]{8mm}{1.0mm}{1.0mm 0.0mm} } } 
    \put(133,120) {$\Sigma_{\bquark}^{-}$}
    \put(133,115) {$\Sigma_{\bquark}^{\ast-}$}
    \put(133,110) {$\mathrm{NR}$}
    \put(133,105) {total}
    \put(125,135){\lhcb}
    \put(40,135){$\decay{\Lambda_{\bquark}\mathrm{(6152)}^0}{\Lb\pip\pim}$}
    \put(40, 71){$\decay{\Lambda_{\bquark}\mathrm{(6146)}^0}{\Lb\pip\pim}$}
    \put(4,107){\begin{sideways}$N_{\Lambda_{\bquark}\mathrm{(6152)}^0}$/(5\mev)\end{sideways}}
    \put(4, 47){\begin{sideways}$N_{\Lambda_{\bquark}\mathrm{(6146)}^0}$/(5\mev)\end{sideways}}
     \put( 43,  2) {$m_{\Lb\pip}$} \put( 69,  2) {$\left[\!\gev\right]$}
     \put(110,  2) {$m_{\Lb\pim}$} \put(134,  2) {$\left[\!\gev\right]$}
  \end{picture}
  \caption { \small
  Background-subtracted mass distribution of (left)~$\Lb\pip$
  and (right)~$\Lb\pim$ combinations from 
  (top)~\mbox{$\decay{\Lambda_{\bquark}\mathrm{(6152)}^0}{\Lb\pip\pim}$}
  and (bottom)~\mbox{$\decay{\Lambda_{\bquark}\mathrm{(6146)}^0}{\Lb\pip\pim}$~decays}.
  Results of fits with a~model comprising  $\Sigma_{\bquark}$,
  $\Sigma_{\bquark}^{\ast}$ and nonresonant\,(NR) components 
  are superimposed.
  }
  \label{fig:figure3}
\end{figure}
}


To~probe further the~resonance structure of 
the~$\decay{\Lambda_{\bquark}\mathrm{(6146)}^0}{\Lb\pip\pim}$ and 
$\decay{\Lambda_{\bquark}\mathrm{(6152)}^0}{\Lb\pip\pim}$~decays, 
the~background\nobreakdash-subtracted $\Lb\Ppi^{\pm}$~mass spectra are studied.
The~{\sc{sPlot}}~technique~\cite{Pivk:2004ty} is used here; 
it  projects out the~signal components from 
the~combined signal\nobreakdash-plus\nobreakdash-background densities
using $m_{\Lb\pip\pim}$~as a~discriminating variable.
The~resulting $\Lb\Ppi^{\pm}$~mass spectra
are shown in Fig.~\ref{fig:figure3}.
The~spectra are fit with three components,
describing the~contributions from $\Sigma_{\bquark}^{\pm}$,
$\Sigma_{\bquark}^{\ast\pm}$ and nonresonant decays. 
Relativistic S- and P-wave Breit$-$Wigner functions 
are used to describe $\Sigma_{\bquark}^{\pm}\to{\Lb\Ppi^{\pm}}$
and $\Sigma_{\bquark}^{(\ast)\pm}\to{\Lb\Ppi^{\pm}}$~decays, 
respectively. 
The~choice of the~orbital angular 
momentum is based on the~quark model expectation 
of spin~$\frac{1}{2}$ for $\Lb$ and $\Sigma_{\bquark}$~baryons
and~$\frac{3}{2}$ for $\Sigma_{\bquark}^{\ast}$~states.
Since the resolution
on the~$\Lb\Ppi^{\pm}$~mass is much better than
the~natural widths of the~$\Sigma_{\bquark}^{(\ast)\pm}$~states,
resolution effects are neglected.
The~nonresonant component is parameterised as a~product of 
two\nobreakdash-from\nobreakdash-three\nobreakdash-body decay phase space functions~\cite{Byckling}
and a~first\nobreakdash-order polynomial function.
The~masses and widths of the~$\Sigma_{\bquark}^{(\ast)\pm}$~states
are fixed to their~known values~\cite{LHCb-PAPER-2018-032}. 
The results of extended unbinned maximum-likelihood fits to 
the~background\nobreakdash-subtracted $\Lb\Ppi^{\pm}$~mass
distributions are shown in Fig.~\ref{fig:figure3}, 
and are presented in Table~\ref{tab:sigmabs} 
of the~Supplemental Material.
Significant
$\decay{\Lambda_{\bquark}\mathrm{(6152)}^0}{\Sigma_{\bquark}^{\pm}\Ppi^{\mp}}$
and 
$\decay{\Lambda_{\bquark}\mathrm{(6152)}^0}{\Sigma_{\bquark}^{\ast\pm}\Ppi^{\mp}}$~signals
are observed, accounting for approximately  one-third  and one-quarter of 
the~signal decays
in the~sample, respectively.
The~statistical significance of the~contributions
is in excess of seven and five~standard deviations, respectively.
For~the~$\Lambda_{\bquark}\mathrm{(6146)}^0$~state, 
$\decay{\Lambda_{\bquark}\mathrm{(6146)}^0}{\Sigma_{\bquark}^{\ast\pm}\Ppi^{\mp}}$~decays
account for about half of the~observed decay  rate 
with a~statistical significance in excess of six~standard deviations.
No~significant  
$\decay{\Lambda_{\bquark}\mathrm{(6146)}^0}{\Sigma_{\bquark}^{\pm}\Ppi^{\mp}}$~signals
are observed.

Several sources of systematic uncertainty are considered.
The most important source of systematic uncertainty on the mass measurements 
derives from the~knowledge of the momentum scale.
This uncertainty is evaluated by varying the momentum scale 
within its known~uncertainty~\cite{LHCb-PAPER-2013-011} 
and rerunning the mass fit. 
The~second uncertainty arises from the assumed parameters of the~Breit$-$Wigner functions.
To~estimate this uncertainty,
the~orbital angular momentum is changed from $L=0$ to~$2$ for all signal components 
and the~Blatt$-$Weisskopf breakup radii are varied from 1.5 to $5\gev^{-1}$.
Since the~states are narrow and far from the~thresholds, 
the~fitted masses and widths have only very small 
dependency on the assumed parameters. 
The~maximal changes to the~fitted parameters with respect to 
the~baseline fit 
are assigned as systematic uncertainties.
The~impact of the~background model is evaluated by varying the order 
of the polynomial functions from two to four.
A~further source of uncertainty on the~determination of the~natural 
widths arises from known differences in resolution between data and simulation. 
This~effect is assessed by varying conservatively the width of 
the~resolution function by $\pm 10$\%, based on previous studies~\cite{LHCb-PAPER-2014-002,LHCb-PAPER-2014-061,LHCb-PAPER-2017-002,LHCb-PAPER-2017-036,LHCb-PAPER-2018-013,LHCb-PAPER-2019-005}.

\begin{table}[t]
\caption{\small Summary of the systematic uncertainties for 
the~masses, $m$, and 
widths, $\Gamma$,  of the~$\Lambda_{\bquark}\mathrm{(6146)}^0$ and 
$\Lambda_{\bquark}\mathrm{(6152)}^0$ states.
All~values are in \kev.}
\centering
\vspace{0.5mm}
\begin{tabular*}{0.99\linewidth}{@{\hspace{1mm}}l@{\extracolsep{\fill}}cccc@{\hspace{1mm}}}
\multirow{2}{*}{Source}
& \multicolumn{2}{c}{$\Lambda_{\bquark}\mathrm{(6146)}^0$}
& \multicolumn{2}{c}{$\Lambda_{\bquark}\mathrm{(6152)}^0$}
\\
       & $m$ 
       & $\Gamma$ 
       & $m$ 
       & $\Gamma$
       \\
\hline
Momentum scale   
& \phantom{0}80 &       ---         
& $\phantom{-}80$ &      ---          \\
Signal model     
& \phantom{0}50 & \phantom{0}50 
&  $\phantom{0}50$ & \phantom{0}50  \\
Resolution model 
&  \phantom{0}15 &          270   
&    $\!\!\!\!<10$       &           310  \\
Background model 
& \phantom{0}30 & \phantom{0}30 
& $\phantom{0}30$ & \phantom{0}20  \\ 
\hline 
Total            
& 100 & 280 
& 100 & 320 \\
Including $\Lb$~mass systematic 
& 220 & 280 
& 220 & 320
\end{tabular*}
\label{tab:sys}
\end{table}

The different sources of systematic uncertainty 
are summarised in Table~\ref{tab:sys}. In~all cases they are smaller than 
the~statistical uncertainties.
A~large part of the~systematic uncertainty cancels for 
the mass~splitting, $\Delta m$,
between the~$\Lambda_{\bquark}\mathrm{(6146)}^0$
and $\Lambda_{\bquark}\mathrm{(6152)}^0$~states. 
The remaining systematic uncertainty for $\Delta m$~is~20\kev.
An~additional uncertainty arises due to 
the~value of the~\Lb~mass used in the~constrained fit.
The~statistical uncertainty on the~\Lb~mass introduces 
an~uncertainty of 0.16\mev on the~$\Lambda_{\bquark}\mathrm{(6146)}^0$ and 
$\Lambda_{\bquark}\mathrm{(6152)}^0$masses. 
This~uncertainty is quoted separately.  
The~systematic uncertainty on the~constraint is correlated, 
through the~momentum scale, with the~masses measured in this 
analysis and is instead included in the~final 
systematic uncertainty in Table~\ref{tab:sys}.

In summary, a~new structure with high statistical significance 
is observed in the~$\Lb\pip\pim$~mass spectrum 
using $\decay{\Lb}{\Lc\pim}$~decays,
and confirmed using a~sample of $\Lb$~baryons
reconstructed through the~$\decay{\Lb}{\jpsi\proton\Km}$~decay.
An~analysis of the~$\Lb\pip\pim$~mass spectra for the~regions
enriched by the~$\Sigma_{\bquark}^{(\ast)\pm}$~resonances 
suggests the~interpretation of the~structure as  two almost degenerate 
narrow states, denoted as $\Lambda_{\bquark}\mathrm{(6146)}^0$
and  $\Lambda_{\bquark}\mathrm{(6152)}^0$.
The~masses and natural  widths of these states are measured to be 
\begin{eqnarray*}
    m_{\Lambda_{\bquark}\mathrm{(6146)}^0}      & = &  6146.17 \pm 0.33 \pm0.22 \pm 0.16 \mev \,, \\
    m_{\Lambda_{\bquark}\mathrm{(6152)}^0}      & = &  6152.51 \pm 0.26 \pm0.22 \pm 0.16 \mev \,, \\
   \Gamma_{\Lambda_{\bquark}\mathrm{(6146)}^0} & = &  \phantom{000}2.9\phantom{0} \pm1.3\phantom{0}\pm0.3\phantom{0}\mev \,, \\ 
   \Gamma_{\Lambda_{\bquark}\mathrm{(6152)}^0} & = &  \phantom{000}2.1\phantom{0} \pm 0.8\phantom{0} \pm0.3 \phantom{0}\mev \,, 
\end{eqnarray*}
where the first uncertainty is statistical, the second systematic and 
the third for the~mass measurements
due to imprecise knowledge of the~mass of the~$\Lb$~baryon.
The~mass differences with respect to the~\Lb~mass 
are measured to be 
\begin{eqnarray*}
   m_{\Lambda_{\bquark}\mathrm{(6146)}^0} - m_{\Lb} & = &  526.55 \pm 0.33 \pm0.10\mev \,, \\
   m_{\Lambda_{\bquark}\mathrm{(6152)}^0} - m_{\Lb} & = &  532.89 \pm 0.26 \pm0.10\mev \,,  
\end{eqnarray*}
and the~mass difference between the two states is measured 
to be \mbox{$6.34 \pm 0.32 \pm0.02 \mev$}. 
The masses of the two states measured in this analysis are consistent with
the~predictions for the~doublet of $\Lambda_{\bquark}\mathrm{(1D)}^0$~states 
with quantum numbers\,(spin $\mathrm{J}$~and parity~$\mathrm{P}$) 
$\mathrm{J^P}=\tfrac{3}{2}^+$ and 
$\tfrac{5}{2}^+$~\cite{Chen:2014nyo, Capstick:1986bm}. 
Similar natural widths
are expected for the~two states of the~doublet in HQET~\cite{Isgur:1991wq}.
The~observed decay pattern, where one of the states decays to both $\Sigma_{\bquark}$
with $\mathrm{J^P}=\tfrac{1}{2}^+$ and $\Sigma_{\bquark}^{\ast}$ 
with $\mathrm{J^P}=\tfrac{3}{2}^+$, while the other decays primarily to 
$\Sigma_{\bquark}^{\ast}$, is also consistent  with the~above assignment. 
However, the~interpretation of these states as excited $\Sigma_{\bquark}^0$
states cannot be excluded.

\section*{Acknowledgements}
%
%
\noindent We express our gratitude to our colleagues in the~CERN
accelerator departments for the excellent performance of the LHC. We
thank the technical and administrative staff at the LHCb
institutes.
We acknowledge support from CERN and from the national agencies:
CAPES, CNPq, FAPERJ and FINEP\,(Brazil); 
MOST and NSFC\,(China); 
CNRS/IN2P3\,(France); 
BMBF, DFG and MPG\,(Germany); 
INFN\,(Italy); 
NWO\,(Netherlands); 
MNiSW and NCN\,(Poland); 
MEN/IFA\,(Romania); 
MSHE\,(Russia); 
MinECo\,(Spain); 
SNSF and SER\,(Switzerland); 
NASU\,(Ukraine); 
STFC\,(United Kingdom); 
DOE NP and NSF\,(USA).
We~acknowledge the computing resources that are provided by CERN, 
IN2P3\,(France), 
KIT and DESY\,(Germany), 
INFN\,(Italy), 
SURF\,(Netherlands),
PIC\,(Spain), 
GridPP\,(United Kingdom), 
RRCKI and Yandex LLC\,(Russia), 
CSCS\,(Switzerland), 
IFIN\nobreakdash-HH\,(Romania), 
CBPF\,(Brazil),
PL\nobreakdash-GRID\,(Poland) and OSC\,(USA).
We~are indebted to the communities behind the multiple open-source
software packages on which we depend.
Individual groups or members have received support from
AvH Foundation\,(Germany);
EPLANET, Marie Sk\l{}odowska\nobreakdash-Curie Actions and ERC\,(European Union);
ANR, Labex P2IO and OCEVU, and R\'{e}gion Auvergne\nobreakdash-Rh\^{o}ne\nobreakdash-Alpes\,(France);
Key Research Program of Frontier Sciences of CAS, CAS PIFI, and the Thousand Talents Program\,(China);
RFBR, RSF and Yandex LLC\,(Russia);
GVA, XuntaGal and GENCAT\,(Spain);
the Royal Society
and the~Leverhulme Trust\,(United Kingdom).





\clearpage
\addcontentsline{toc}{section}{References}
\setboolean{inbibliography}{true}
\bibliographystyle{LHCb}
\bibliography{standard,local,LHCb-PAPER,LHCb-CONF,LHCb-DP,LHCb-TDR}

\newpage

\centerline
{\large\bf LHCb collaboration}
\begin
{flushleft}
\small
R.~Aaij$^{30}$,
C.~Abell{\'a}n~Beteta$^{47}$,
T.~Ackernley$^{57}$,
B.~Adeva$^{44}$,
M.~Adinolfi$^{51}$,
H.~Afsharnia$^{8}$,
C.A.~Aidala$^{78}$,
S.~Aiola$^{24}$,
Z.~Ajaltouni$^{8}$,
S.~Akar$^{62}$
% ...

{\footnotesize \it

% ...
$ ^{8}$Universit{\'e} Clermont Auvergne, CNRS/IN2P3, LPC, Clermont-Ferrand, France\\
% ...
$ ^{24}$INFN Sezione di Milano, Milano, Italy\\
% ...
$ ^{30}$Nikhef National Institute for Subatomic Physics, Amsterdam, Netherlands\\
% ...
$ ^{44}$Instituto Galego de F{\'\i}sica de Altas Enerx{\'\i}as (IGFAE), Universidade de Santiago de Compostela, Santiago de Compostela, Spain\\
% ...
$ ^{47}$Physik-Institut, Universit{\"a}t Z{\"u}rich, Z{\"u}rich, Switzerland\\
% ...
$ ^{51}$H.H. Wills Physics Laboratory, University of Bristol, Bristol, United Kingdom\\
% ...
$ ^{57}$Oliver Lodge Laboratory, University of Liverpool, Liverpool, United Kingdom\\
% ...
$ ^{62}$University of Cincinnati, Cincinnati, OH, United States\\
% ...
$ ^{65}$Syracuse University, Syracuse, NY, United States\\
% ...
$ ^{78}$University of Michigan, Ann Arbor, United States, associated to $^{65}$\\
\bigskip
% ...
}
\end{flushleft}

\clearpage
\renewcommand{\thetable}{S\arabic{table}}   
\renewcommand{\thefigure}{S\arabic{figure}}
\setcounter{figure}{0}
\setcounter{table}{0}
\ifthenelse{\boolean{prl}}{
\section{Observation of new resonances 
in the~\boldmath{$\Lb\pip\pim$} system}
}{\section*{Observation of new resonances 
in the~\boldmath{$\Lb\pip\pim$} system}
}
\centerline{\it{Supplemental Material}}
\vspace*{2mm}


\subsection*{The \boldmath{$\decay{\Lb}{\Lc\pim}$}
and \boldmath{$\decay{\Lb}{\jpsi\proton\Km}$}~candidates}

The~mass distributions for selected $\decay{\Lb}{\Lc\pim}$
and $\decay{\Lb}{\jpsi\proton\Km}$~candidates are shown in 
Fig.~\ref{fig:Lb}.  
The~distributions are fit with a sum of 
a~signal and a background component. 
The~signal component is parameterised 
by a~modified Gaussian function with power\nobreakdash-law 
tails on both sides of the~peak,
\begin{equation*}
  G\left(m;\Pmu,\sigma,
  \Palpha_{\mathrm{L}},
  \Palpha_{\mathrm{R}},n_{\mathrm{L}},n_{\mathrm{R}}\right)
  \propto
  \left\{
  \begin{array}{lcc}
    A_{\mathrm{L}} \left( \frac{n_{\mathrm{L}}+1}{ n_{\mathrm{L}}+1 - \Palpha_{\mathrm{L}} (  \Palpha_{\mathrm{L}} + \Pdelta m  )} \right)^{n_{\mathrm{L}}+1}  & \mathrm{for}  &   \Pdelta m < -\Palpha_{\mathrm{L}}, \\
    \dfrac{1}{\sqrt{2\pi}\sigma}\mathrm{e}^{-\frac{1}{2}\Pdelta m^2} &  \mathrm{for} &    -\Palpha_{\mathrm{L}} < \Pdelta m < \Palpha_{\mathrm{R}}, \\ 
    A_{\mathrm{R}} \left( \frac{ n_{\mathrm{R}}+1} { n_{\mathrm{R}}+1 - \Palpha_{\mathrm{R}} (  \Palpha_{\mathrm{R}} - \Pdelta m ) } \right)^{n_{\mathrm{R}}+1}  & \mathrm{for}  &   \Pdelta m > \phantom{-}\Palpha_{\mathrm{R}}, 
  \end{array}
  \right.
\end{equation*}
where $\Pdelta m \equiv \tfrac{ m - \Pmu}{\sigma}$,
$n_{{\mathrm{L}},{\mathrm{R}}}>0$, $\Palpha_{\mathrm{L},\mathrm{R}}>0$ and $A_{\mathrm{L},\mathrm{R}}= \tfrac{1}{\sqrt{2\pi}\sigma}\mathrm{e}^{-\frac{1}{2}\alpha_{\mathrm{L},\mathrm{R}}^2}$.
The~background is parameterised by 
the~product 
of an~exponential function and a second\nobreakdash-order polynomial function.
The~signal yields  
are listed in Table~\ref{tab:Lb}.

\ifthenelse{\boolean{prl}}{
\begin{figure}[bh]
  \setlength{\unitlength}{1mm}
  \centering
   \includegraphics*[height=140mm,
    ]{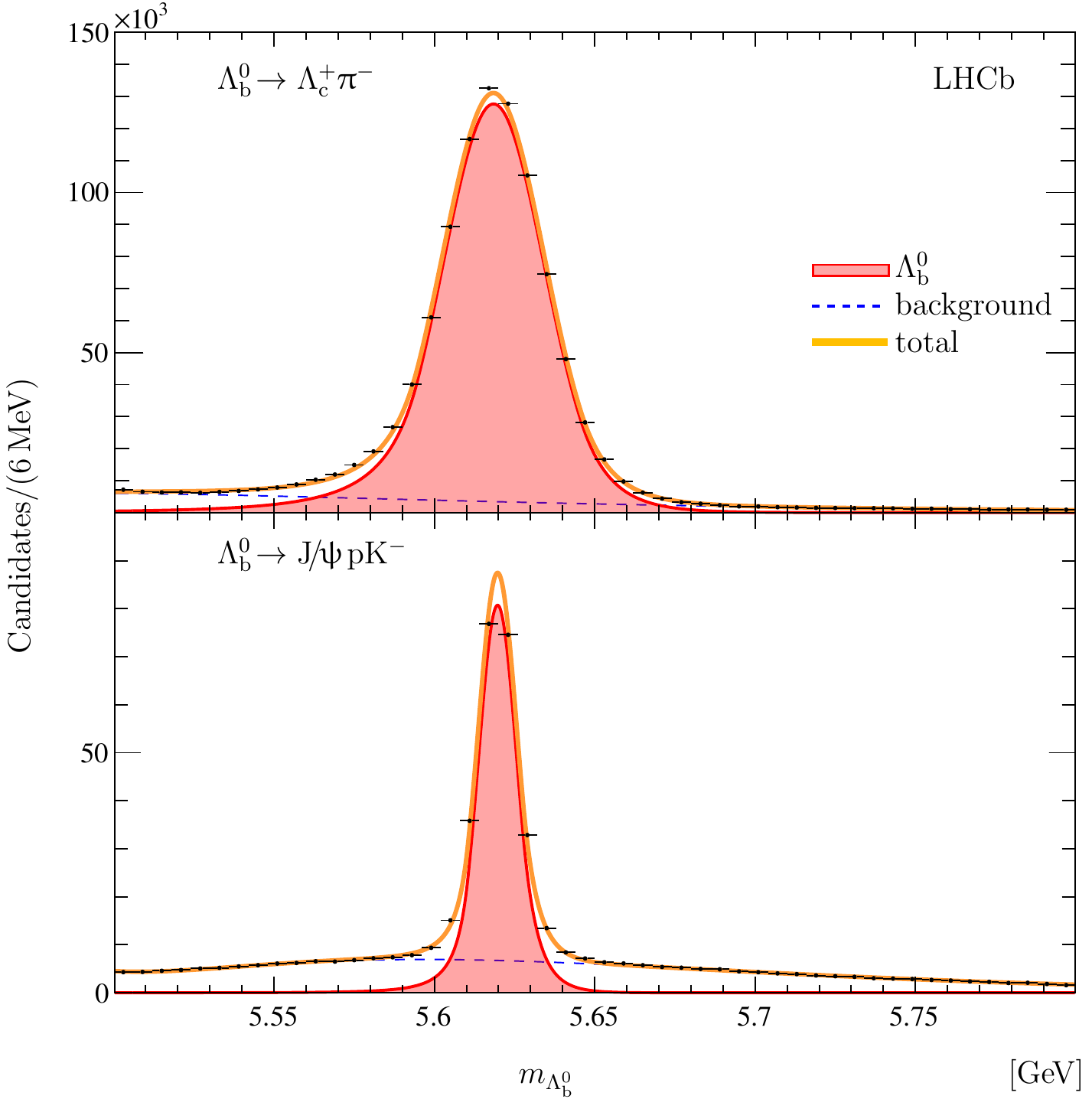}
  \caption { \small
    Mass distribution  of selected $\Lb$~candidates
  from the (top)~$\decay{\Lb}{\Lc\pim}$
  and (bottom)~$\decay{\Lb}{\jpsi\proton\Km}$ decay modes.
  }
  \label{fig:Lb}
\end{figure}
}{
\begin{figure}[htb]
  \setlength{\unitlength}{1mm}
  \centering
  \begin{picture}(150,150)
    %
    \put(  0, 0){ 
      \includegraphics*[width=150mm,height=150mm,%
       ]{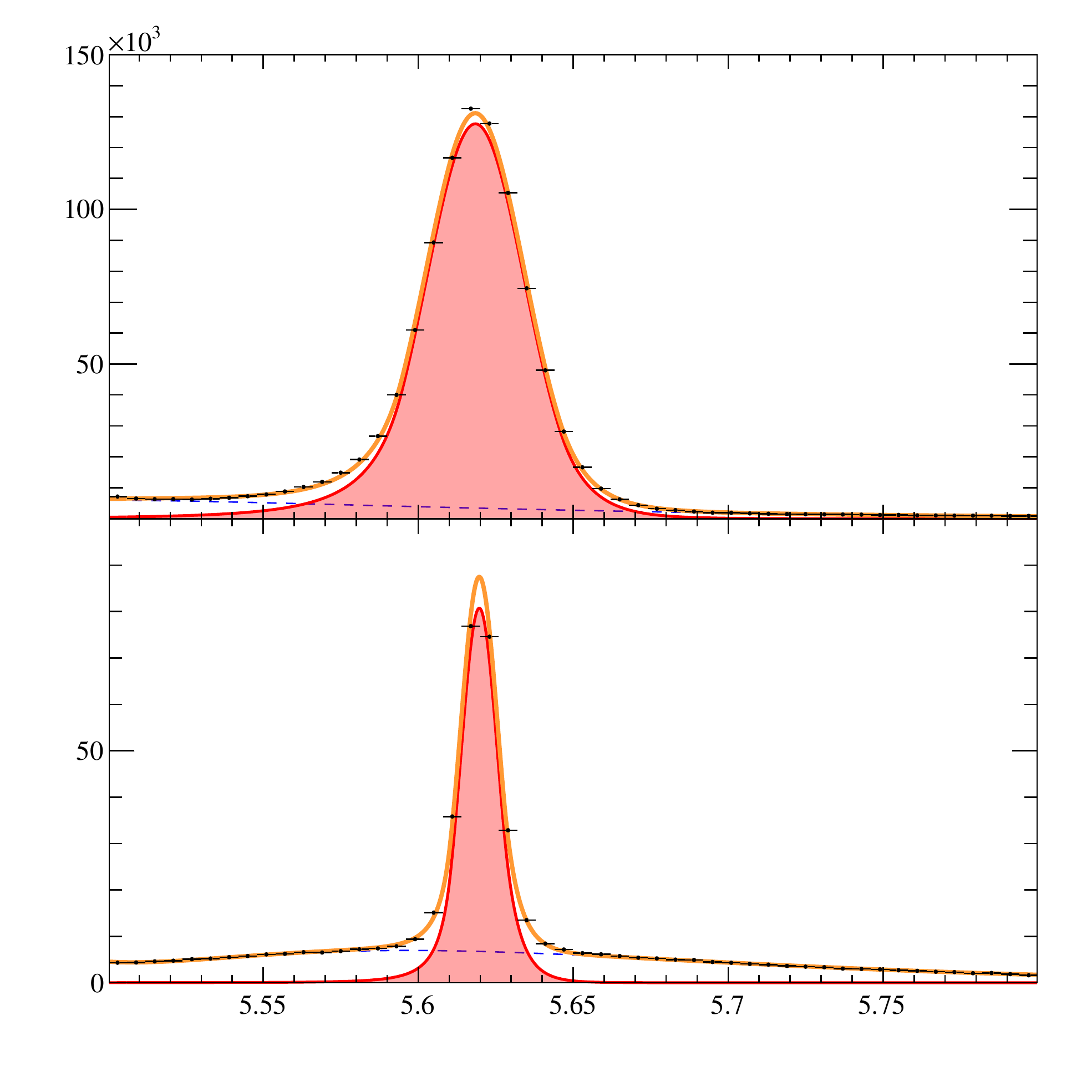}
    }
    \put(109,110) {\begin{tikzpicture}[x=1mm,y=1mm]\filldraw[fill=red!35!white,draw=red,thick]  (0,0) rectangle (10,1.5);\end{tikzpicture} }
    \put(109,105) {\color[rgb]{0.00,0.00,1.00} {\hdashrule[0.5ex][x]{10mm}{0.5mm}{1.0mm 1.0mm} } } 
    \put(109,100) {\color[rgb]{1.00,0.75,0.00} {\hdashrule[0.5ex][x]{10mm}{1.0mm}{1.0mm 0.0mm} } } 
    \put(120,110) {$\Lb$}
    \put(120,105) {$\mathrm{background}$}
    \put(120,100) {total}
    \put(125,135){\lhcb}
   \put(2,60){\begin{sideways}Candidates/(6\mev)\end{sideways}}
    \put(30,135){$\decay{\Lb}{\Lc\pim}$}
    \put(30, 72){$\decay{\Lb}{\jpsi\proton\Km}$}
  \put( 70,  3) {$m_{\Lb}$} \put( 135,  3) {$\left[\!\gev\right]$}
  \end{picture}
  \caption { \small
    Mass distribution  of selected $\Lb$~candidates
  from the (top)~$\decay{\Lb}{\Lc\pim}$
  and (bottom)~$\decay{\Lb}{\jpsi\proton\Km}$ decay modes.
  }
  \label{fig:Lb}
\end{figure}
}

\begin{table}[htb]
  \centering
  \caption{\small
    The signal yields for 
    $\decay{\Lb}{\Lc\pim}$~and
    $\decay{\Lb}{\jpsi\proton\Km}$~decays.
  }\label{tab:Lb}
  \vspace*{0.5mm}
  \begin{tabular*}{0.45\linewidth}{@{\hspace{3mm}}l@{\extracolsep{\fill}}c@{\hspace{3mm}}}
    Decay mode
    & $N~\left[10^3\right]$
    \\[1mm]
    \hline 
    \\[-2mm]
 $\decay{\Lb}{\Lc\pim}$
 & $892.8\pm1.2$  
 \\
  $\decay{\Lb}{\jpsi\proton\Km}$
 & $217.5\pm0.7$
   \end{tabular*}
\end{table}

\subsection*{Results of the simultaneus  fit to 
\boldmath{$\Lb\pip\pim$}~mass spectra 
in the~three $\Lb\Ppi^{\pm}$~mass regions}

The yields of the~$\Lambda_{\bquark}\mathrm{(6146)}^0$ 
and  $\Lambda_{\bquark}\mathrm{(6152)}^0$~signals 
from  the simultaneous extended unbinned maximum-likelihood 
fit to  the~$\Lb\pip\pim$~mass spectra in 
the~three $\Lb\Ppi^{\pm}$~mass regions are 
presented in Table.~\ref{tab:Lbstars}.

\begin{table}[htb]
  \centering
  \caption{\small
  The~yields of 
  the~$\Lambda_{\bquark}\mathrm{(6146)}^0$ 
  and  $\Lambda_{\bquark}\mathrm{(6152)}^0$~signals
  in the three $\Lb\Ppi^{\pm}$~mass regions.
  }\label{tab:Lbstars}
  \vspace*{3mm}
  \begin{tabular*}{0.65\textwidth}{@{\hspace{3mm}}l@{\extracolsep{\fill}}lccc@{\hspace{3mm}}}
    &   $\Sigma_{\bquark}$~region
    &   $\Sigma_{\bquark}^{\ast}$~region
    &   NR~region
    \\[1mm]
    \hline 
    \\[-2mm]
 $N_{\Lambda_{\bquark}\mathrm{(6146)}^0}$
 &  $\phantom{0}67\pm40\phantom{0}$ 
 &  $460\pm92\phantom{0}$ 
 &  $624\pm136$
 \\
 $N_{\Lambda_{\bquark}\mathrm{(6152)}^0}$
 &  $357\pm52\phantom{0}$
 &  $305\pm70\phantom{0}$
 &  $510\pm109$
  \end{tabular*}
\end{table}

\subsection*{Results of the fits to  background-subtracted \boldmath{$\Lb\Ppi^{\pm}$}~mass spectra}

   The yield of $\decay{\Lambda_{\bquark}\mathrm{(6146)}^0}{\Sigma_{\bquark}^{(\ast)\pm}\Ppi^{\mp}}$ and $\decay{\Lambda_{\bquark}\mathrm{(6152)}^0}{\Sigma_{\bquark}^{(\ast)\pm}\Ppi^{\mp}}$~decays, determined from fits to the~background\nobreakdash-subtracted $\Lb\Ppi^{\pm}$~mass distributions, are summarized in Table~\ref{tab:sigmabs}.
   
\begin{table}[htb]
  \centering
  \caption{\small
  The~yields, $N$,  and statistical significance, ${\mathcal{S}}_{\mathrm{W}}$, 
  of the~$\decay{\Lambda_{\bquark}\mathrm{(6146)^0}}{\Sigma_{\bquark}^{(\ast)\pm}\Ppi^{\mp}}$
  and 
  $\decay{\Lambda_{\bquark}\mathrm{(6152)}^0}{\Sigma_{\bquark}^{(\ast)\pm}\Ppi^{\mp}}$~signals
   from the fits to 
   the~background-subtracted $\Lb\Ppi^{\pm}$~mass distributions.
  }\label{tab:sigmabs}
  \vspace*{3mm}
  \begin{tabular*}{0.55\textwidth}{@{\hspace{3mm}}l@{\extracolsep{\fill}}cc@{\hspace{3mm}}}
      &  $N$  & $\mathcal{S}_{\mathrm{W}}$
      \\[1mm]
    \hline 
    \\[-2mm]
   $\decay{\Lambda_{\bquark}\mathrm{(6152)}^0}{\Sigma_{\bquark}^{+}\pim}$
   & $213\pm44$ & $7.8\sigma$
   \\
  $\decay{\Lambda_{\bquark}\mathrm{(6152)}^0}{\Sigma_{\bquark}^{-}\pip}$
   & $208\pm43$ & $7.6\sigma$ 
   \\
  $\decay{\Lambda_{\bquark}\mathrm{(6152)}^0}{\Sigma_{\bquark}^{\ast+}\pim}$
   & $163\pm45$ & $5.3\sigma$ 
   \\
   $\decay{\Lambda_{\bquark}\mathrm{(6152)}^0}{\Sigma_{\bquark}^{\ast-}\pip}$
   & $141\pm45$ & $4.5\sigma$
   \\
   $\decay{\Lambda_{\bquark}\mathrm{(6146)}^0}{\Sigma_{\bquark}^{+}\pim}$
   & $\phantom{0}53\pm30$ & $2.3\sigma$  
   \\
  $\decay{\Lambda_{\bquark}\mathrm{(6146)}^0}{\Sigma_{\bquark}^{-}\pip}$
   & $\phantom{00}0\pm20$ & ---
   \\
   $\decay{\Lambda_{\bquark}\mathrm{(6146)}^0}{\Sigma_{\bquark}^{\ast+}\pim}$
   & $285\pm51$ & $8.4\sigma$
   \\
   $\decay{\Lambda_{\bquark}\mathrm{(6146)}^0}{\Sigma_{\bquark}^{\ast-}\pip}$
   & $227\pm52$ & $6.3\sigma$  
  \end{tabular*}
\end{table}

\end{document}